\documentclass[twoside,twocolumn]{article}

\usepackage{textcomp}

\usepackage{xspace}

\usepackage{gensymb}
\usepackage{float}
\usepackage{multicol}
\usepackage{physics}
\usepackage{siunitx}
\usepackage{amsfonts}
\usepackage{amssymb}
\usepackage{verbatim}
\usepackage{hyperref}
\usepackage{url} 
\usepackage{graphicx}
\usepackage{newtxtext,newtxmath}
\usepackage{xcolor}
\usepackage{bm}
\newcommand{\credval}[4]{\ensuremath{#1 = \displaystyle #2^{+#3}_{-#4}}}

\usepackage{mathtools}
\usepackage[citestyle=numeric-comp,bibstyle=numeric, sorting=none, natbib=true, backend=bibtex,hyperref=true,maxbibnames=5,maxcitenames=2]{biblatex}
\usepackage{blindtext} 

\usepackage[english]{babel} 

\usepackage[hmarginratio=1:1, left = 18mm, top=32mm,columnsep=20pt]{geometry} 
\usepackage[hang, small,labelfont=bf,up,textfont=it,up]{caption} 
\usepackage{booktabs} 

\usepackage{abstract} 

\usepackage{titlesec} 

\def\gt#1{Georgia Institute of Technology#1 (GeorgiaTech#1)\gdef\gt{GeorgiaTech}}
\def\si#1{Senior Investigator#1 (SI#1)\gdef\si{SI}}
\def\emri#1{Extreme Mass-Ratio Inspiral#1 (EMRI#1)\gdef\emri{EMRI}}
\def\imbh#1{Intermediate Mass Black Hole#1 (IMBH#1)\gdef\imbh{IMBH}}
\def\smbh#1{supermassive black hole#1(SMBH#1)\gdef\smbh{SMBH}}
\def\bbh#1{binary black hole#1 (BBH#1)\gdef\bbh{BBH}}
\def\pbh#1{primordial black hole#1 (PBH#1)\gdef\pbh{PBH}}
\def\imbhb#1{intermediate mass black hole binary#1 (IMBHB#1)\gdef\imbhb{IMBHB}}
\def\hmns#1{hypermassive neutron star#1 (HMNS#1)\gdef\hmns{HMNS}}
\def\bh#1{black hole#1 (BH#1)\gdef\bh{BH}}
\def\ns#1{neutron star#1 (NS#1)\gdef\ns{NS}}
\def\hmns#1{hyper-massive neutron star#1 (HMNS#1)\gdef\hmns{HMNS}}
\def\nsbh#1{neutron star-black hole#1 (NSBH#1)\gdef\nsbh{NSBH}}
\def\bns#1{binary neutron star#1 (BNS#1)\gdef\bns{BNS}}
\def\gw#1{gravitational wave#1 (GW#1)\gdef\gw{GW}}
\def\eos#1{equation of state#1 (EOS#1)\gdef\eos{EOS}}
\def\gpu#1{graphics processing unit#1 (GPU#1)\gdef\gpu{GPU}}
\def\gr#1{General Relativity#1 (GR#1)\gdef\gr{GR}}
\def\cbc#1{Compact Binary Coalescence#1 (CBC#1)\gdef\cbc{CBC}}
\def\eob#1{Effective-One-Body#1 (EOB#1)\gdef\eob{EOB}}
\def\pnw#1{post-Newtonian#1 (PN#1)\gdef\pnw{PN}}
\def\pmw#1{post-Minkowskian#1 (PM#1)\gdef\pmw{PM}}
\def\hom#1{Higher Order Mode#1 (HOM#1)\gdef\hom{HOM}}
\def\agn#1{Active Galactic Nuclei#1 (AGN#1)\gdef\agn{AGN}}

\def\ligo#1{Laser Interferometer Gravitational-Wave Observatory#1 (LIGO#1)\gdef\ligo{LIGO}}
\def\lisa#1{Laser Interferometer Space Antennae#1 (LISA#1)\gdef\lisa{LISA}}

\renewcommand\thesection{\Roman{section}} 
\renewcommand\thesubsection{\roman{subsection}} 
\titleformat{\section}[block]{\large\scshape\centering}{\thesection.}{1em}{} 
\titleformat{\subsection}[block]{\large}{\thesubsection.}{1em}{} 

\usepackage{titling} 

\usepackage{hyperref} 

\usepackage[normalem]{ulem}
\usepackage{soul}   

\title{Parameter estimation of gravitational waves from hyperbolic black hole encounters} 
\author{
Chad Henshaw\textsuperscript{1},
Jacob Lange\textsuperscript{3,4},
Peter Lott\textsuperscript{1,}\textsuperscript{2},
Richard O'Shaughnessy\textsuperscript{5},
Laura Cadonati\textsuperscript{1} \\[1ex]
\small
\textsuperscript{1} Center for Relativistic Astrophysics, Georgia Institute of Technology, Atlanta, GA 30332, USA\\
\small
\textsuperscript{2} Phenikaa Institute for Advanced Study, Phenikaa University, Hanoi, Vietnam\\
\small
\textsuperscript{3} Istituto Nazionale di Fisica Nucleare - Sezione di Torino, Torino, Italy\\
\small
\textsuperscript{4} Center of Gravitational Physics, University of Texas at Austin, Austin, TX 78712, USA\\
\small
\textsuperscript{5} Center for Computational Relativity and Gravitation, Rochester Institute of Technology, Rochester, NY 14623, USA
}
\date{\today} 
\renewcommand{\maketitlehookd}{%
\begin{abstract}
\noindent Systems of two black holes with unbound orbits can produce a diverse array of gravitational wave signals with rich morphology. This parameter space encompasses both hyperbolic orbit scattering events and dynamical captures, including zoom-whirl orbits with multiple flybys and direct plunge mergers. These signals challenge traditional parameter estimation infrastructure, which is largely optimized for quasicircular inspiral binaries. In this work we discuss the adaptation of the Rapid Iterative FiTting (RIFT) algorithm to this problem using the \textit{TEOBResumSDALI} waveform model which can simulate generic orbits. We present results from a study of simulated signals emulating a scatter and plunge event, utilizing the design sensitivity of the forthcoming Cosmic Explorer interferometer. Our analysis demonstrates that RIFT accurately recovers the mass, spins, and hyperbolic orbit parameters: the system energy and angular momentum defined at a fiducial initial separation.

\end{abstract}
}

\begin{document}

\maketitle

\textit{Introduction} - Over the last ten years the LIGO-Virgo-KAGRA (LVK) collaboration have detected over 200 gravitational wave (GW) events from compact binary coalescences (CBCs)~\cite{gwtc1, gwtc2, gwtc2.1, gwtc3, aps2025session}, which are predominantly mergers of black hole systems with quasicircular orbits. However, many more GW sources are expected to exist and are actively searched for by the LVK~\cite{Abbott2021_short, Abbott2021_long}. In regions dense with stellar remnants such as globular clusters~\cite{Dymnikova1982, Heggie1996, Oleary2006, Kocsis2006, 2008MPLA...23...99C, Chatterjee2017}, galactic nuclei~\cite{Dymnikova1982, OLeary2009}, AGN disks~\cite{Trani2019a, Trani2019b, Lott2024}, and clusters of primordial black holes in galactic halos~\cite{Clesse2015, Garcia-Bellido2017, Garcia-Bellido2021}, it is expected that unbound pairs of black holes can make close hyperbolic encounters. These rapid interactions in which the pair either scatters or becomes bound can produce bursts of GW~\cite{Turner1977, Turner1978, Kovacs1977, Kovacs1978, Capozziello2008, DeVittori2012, Damour2014, DeVittori2014, Cho2018, Grobner2020, Nagar2020, Damour2022}. GW from these encounters give potentially numerous probes into these environments. Unlike the quasi-periodic CBC signals, hyperbolic encounters generate short-lived, single-cycle or few-cycle, broadband signals\footnote{Note that high mass CBC signals can resemble a dynamical capture - see e.g.~\cite{Gamba2023}}. Similar to the CBC case, they encode information on the dynamics of their originating system. Accurate parameter estimation (PE) of such signals can extract details about eccentricity, impact parameter, velocity, and scattering angle offering insights into the population and environmental conditions that lead to such events.\par

Recent surveys suggest that such events could be detected by LIGO, Virgo, and KAGRA if the black holes are sufficiently massive and have close periastron passages. While to date no hyperbolic encounter has been detected in GW data~\cite{Morras:2021atg, Bini:2023gaj}, prospects for detection increase significantly~\cite{Kocsis2006, Mukherjee2021} with the advent of proposed third-generation detectors like Cosmic Explorer ~\cite{Abbott2017, CE2019, Evans2021, Hall2022} and the Einstein Telescope~\cite{Punturo2010, Hild2011}. There have been significant past developments in simulating the waveforms for such systems - see e.g. ~\cite{Peters1963, Quinlan1987, Quinlan1989, Gold2013, East2013, Damour2014, Damour2016, Damour2017, Damour2018, Cho2018, Cho2022, Nagar2020, Nagar2021a, Nagar2021b, Damour2022, Khalil2022,  Bae2020, Nagar2020, Bae2024, Andrade2024, Nelson2019, Jaraba2021}. However, there has only been limited development of PE infrastructure needed to use hyperbolic waveforms~\cite{Gamba2023, Fontbute2025}. In these studies, the waveforms were either limited to certain configurations or to certain parts of parameter space. To produce PE on the expected scale for these systems, a comprehensive and flexible algorithm is needed to handle the different types of waveform and produce fast and accurate results.\par


As proof of principle, we demonstrate that we can produce accurate PE for multiple types of close hyperbolic encounters with dramatically different signal features. For all the results presented, we used RIFT~\cite{Pankow2015, Lange2018, Wofford2023}, a grid-based PE code that first marginalizes over extrinsic parameters and then iteratively interpolates the marginalized grid in tandem with Bayes' Theorem to produce the posterior. This highly-parallelizable algorithm allows for the use of more physics-rich but slower models without sacrificing runtimes. The GW model we use for our study is the effective-one-body (EOB) model: \textit{TEOBResumSDALI}~\cite{Chiaramello2020, Nagar2020, Nagar2021a, Nagar2021b, Gamba2023, Nagar2024}. The EOB formalism, which has shown to be reliable even for when the field is strong up to merger and ringdown, has been extended to account for the phenomenology of close hyperbolic encounters.\par

\textit{Methods} - The \textit{TEOBResumSDALI} model can produce GW waveforms from close hyperbolic encounters with non-precessing spins. This leads to an intrinsic parameter space - a set of parameters that determine the system dynamics - spanning $\left\{m_1, m_2, E_0/M, p_\phi^0, \chi_{1,z}, \chi_{2,z}\right\}$. In this parameter space, $m_1$,$m_2$ are the two objects' masses, $E_0/M$ is the energy of the system at initial separation $r_0$, $p_\phi^0$ is the angular momentum of the system at $r_0$, and $\chi_{1,z}$,$\chi_{2,z}$ are the two objects' non-precessing spins aligned (or anti-aligned) with the orbital angular momentum. There are also the extrinsic parameters $\left\{d_L, \theta, \delta, t_c, \iota, \phi, \psi\right\}$, parameters involved in determining the spacetime location and orientation of the system relative to an observer, where $d_L$ is the luminosity distance, $\theta$ is the right ascension, $\delta$ is the declination, $t_c$ is the coalescence time, $\iota$ is the inclination relative to the line of sight, $\phi$ is the phase, and $\psi$ is the polarization angle. While we have computed the posterior probability distribution for the full 13-dimensional parameter space, the figures in this letter focus on the recovery of the intrinsic parameters.\par

This parameter space covers three distinct categories of orbital trajectories that can arise from the initial conditions of the system, and thus three distinct categories of waveforms that must be considered. \textbf{Scattering} events occur when the two compact objects approach each other but do not become gravitationally bound and instead scatter off one another, losing energy to gravitational wave bremsstrahlung. In this case the waveform resembles a short burst with a number of peaks. \textbf{Dynamical capture} events occur when the two objects do become gravitationally bound, and engage in one or more flybys before merging. These are often referred to as zoom-whirl orbits. In this case the waveform resembles that of a highly eccentric CBC system, with one or more pre-merger bursts followed by a merger and ringdown stage. Finally \textbf{plunge} events (direct capture) occur when the two objects merge without any form of inspiral; in this case the waveform consists of a sharp plunge followed by a merger and ringdown. We will henceforth refer to these three waveform classes as \emph{scatters}, \emph{captures}, and \emph{plunges} respectively.\par

\begin{figure*}
    \centering
    \includegraphics[width=\textwidth]{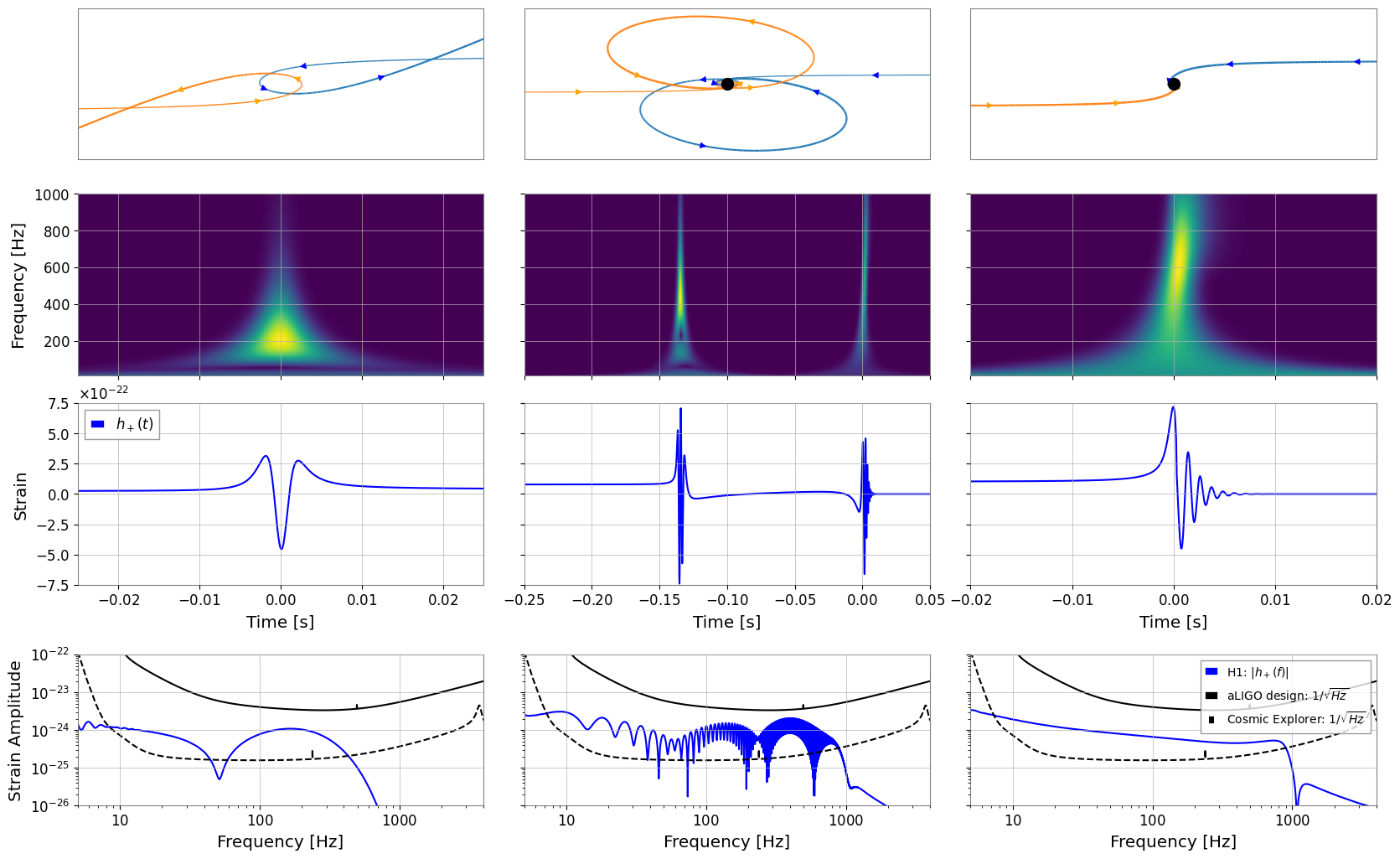}
    \caption{Hyperbolic waveform morphology. The three columns from left to right correspond to the three waveform classes: scatter, dynamical capture, and plunge.The source encounters are simulated at the H1 zenith sky location with total mass $M = 20\, M_\odot$, mass ratio $q = 1$, and zero spin at $dL = 500\,$ Mpc, and hyperbolic parameters $E_0 /M = \left\{1.01, 1.038, 1.051\right\},\: p_\phi^0 = \left\{4.40, 4.70, 4.0\right\}$ respectively. The top row displays the trajectories of the two BHs in the binary's orbital plane, where the blue curve is the path of $m_1$, the orange the path of $m_2$, and the black dot is the final black hole in the capture and plunge cases. The top two rows display the waveform's time-frequency representation~\cite{Henshaw2024} and the strain time series, while the bottom row displays the strain frequency series. The bottom row also includes the design noise curve for the Advanced LIGO detectors as the solid black curve~\cite{KAGRA:2013rdx, Aasi2015} and for Cosmic Explorer as the dashed black curve~\cite{CE2019}.}
    \label{waveform_classes}
\end{figure*}



 A diagram of example waveforms typical of the three classes is presented in Fig.(\ref{waveform_classes}). The diverse nature of the parameter space complicates the implementation of PE, as by nature any parameter estimation engine using this model must be able to sample across a continuous parameter space that contains all three waveform classes. This presents additional complexity, as each waveform class requires different data conditioning. For one, since the majority of data processing happens in the frequency domain, waveform start/ends must be tapered to zero for effective fast Fourier transforms. Scatters have asymmetric pre-event and post-event strain values, while for captures and plunges where there is a merger/ringdown the end strain is always zero, leading to different tapering requirements. Another example is the event/coalescence time $t_c$, which is typically approximated as the time of peak amplitude for the dominant $\left(l,m\right) = \left(2,\pm 2\right)$ mode. However dynamical captures are typified by zoom-whirl orbits with a variable number of pre-merger flybys, and the peak amplitude is often not at the time of merger but at the time of the first flyby. As such for captures the merger time must be located through other means - we use a peak finding algorithm. An effective parameter estimation pipeline for these systems must be able to address these issues; further technical details of our implemented solutions in the RIFT algorithm will be detailed in a forthcoming publication.\par


%

 \textit{Results} - Having overcome these challenges, we will now demonstrate the capability of RIFT to perform PE on hyperbolic systems. Herein we will show the injection and recovery of both a scatter and a plunge waveform. In both cases we simulate signals without adding any additional noise (zero-noise) from systems with total mass $M = m_1 + m_2$ of $20\, M_\odot$, and mass ratio $q \equiv m_2 / m_1, m_1 \geq m_2$ of unity. We assume a Cosmic Explorer detector located at the position of LIGO Hanford.  We orient each system for optimal recovery, with zero inclination at the H1 zenith sky location and luminosity distance of $d_L = 2000\,$ Mpc, yielding SNR of $\sim 42$. We note that a luminosity distance of $d_L = 93\,$ Mpc is needed to achieve similar SNR in a likewise configured network with the design sensitivity of O4 LIGO, highlighting the improvements offered by third-generation detectors. The scatter waveform has hyperbolic parameters $E_0 / M = 1.01,\: p_\phi^0 = 4.40$, and the plunge waveform has hyperbolic parameters $E_0 / M = 1.05,\: p_\phi^0 = 4.00$.\par
 
 For all parameters except $\chi_{1,z},\chi_{2,z}$ and $d_L$, we use a uniform prior distribution across a range of values for a given parameter\footnote{For inclination, we sample uniformly in $\cos\iota$.}. For the masses, we use a uniform prior in the detector frame individual masses $m_1,m_2$ with bounds given in total mass and mass ratio: $M \in \left\{10.0, 200.0\right\} M_\odot,\: q \in \left\{1.0, 10.0\right\}$. The uniform prior range for the hyperbolic parameters are $p_\phi^0 \in \left\{1.0, 10.0\right\}$, with $E_0/M \in \left\{1.0, 1.1\right\}$ for the scatter and $E_0/M \in \left\{1.0, 1.2\right\}$ for the plunge. In both cases, we adopt an aligned-spin prior that is equivalent to the uniform spin magnitude prior after marginalizing out non-aligned spins~\cite{Lange2018}. The prior range for both spins is $\chi_{1,z},\chi_{2,z} \in\{-0.99, 0.99\}$. The distance prior range is $d_L \in \{1.0,10000\}\,$ Mpc and assumes a constant merger rate per unit co-moving volume and cosmological time. Note that in these examples, we evaluate waveforms with only the dominant $\left(l, m\right) = \left(2, \pm 2\right)$ mode from the GW strain mode decomposition.\par

 
 


\begin{figure*}
    \centering
    \includegraphics[width=\textwidth]{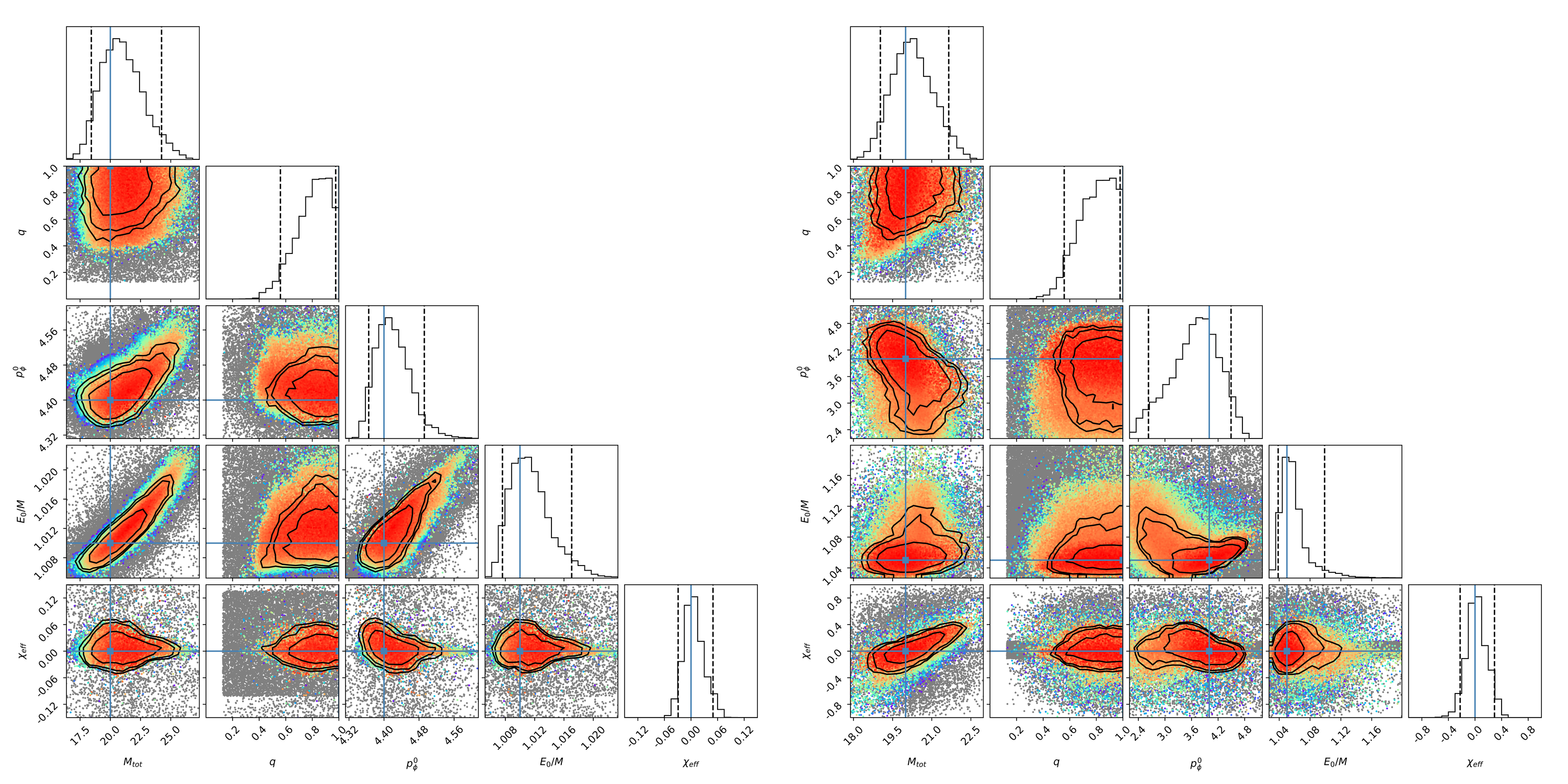}
    \caption{Parameter estimation results for simulated scatter (left panel) and plunge (right panel) signals with zero added noise at SNR$\sim 42$ from a single-detector network at the design sensitivity of Cosmic Explorer~\cite{CE2019}. The 1-D histograms display posterior probability distributions for the parameters $M, q, E_0 / M, p_\phi^0, \chi_{eff}$ and their $5\%$ and $95\%$ percentiles, while the 2-D plots display the distribution of marginalized likelihood in the joint parameter space and contours enclosing the $68\%$, $90\%$, and $95\%$ credible regions. As stated in the text, the SNR for both the scatter and plunge event is ~ 42.}
    \label{combined_corner_free}
\end{figure*}

 
In Fig.(\ref{combined_corner_free}) we see the intrinsic parameter recovery for these two simulated systems. The left-hand panel shows the scatter simulation, and the right-hand panel shows the plunge simulation.  In these corner plots the top diagonal shows the 1-D posterior probability distribution for the per-column parameter, and the remaining plots show the 2-D posterior distribution and the marginalized likelihood $\mathcal{L}$ values across the evaluated grid. The crosshairs denote the injected value for each parameter. The color scale is set from the largest recovered $\ln\left(\mathcal{L}_{max}\right)$ value to a cutoff value of $\ln\left(\mathcal{L}_{max}\right) - 5.0$ for the scatter and $\ln\left(\mathcal{L}_{max}\right) - 20.0$ for the plunge; grid points with marginalized likelihood below this threshold are rendered in gray. The vertical dashed lines on the 1-D histograms denote the $5\%$ and $95\%$ percentiles, and the solid contour lines on the 2-D distribution plots enclose the $68\%$, $90\%$, and $95\%$ credible regions.\par 

















We find that in both cases the total mass distribution is well localized about the true value, yielding recovered median values of \credval{M}{20.89}{3.35}{2.46} $M_\odot$ and \credval{M}{20.24}{1.41}{1.21} $M_\odot$ for the scatter and plunge cases respectively. This constitutes 90\% posterior probability regions localized to just $\sim$3\% and $\sim$1\% of the $M$ prior range respectively. One can see that in the scatter case the hyperbolic parameters are also very well recovered, yielding median values of \credval{E_0/M}{1.0111}{0.0059}{0.0035} and \credval{p_\phi^0}{4.417}{0.075}{0.052}, spanning $\sim$9\% and $\sim$1\% of the $E_0/M$ and $p_\phi^0$ prior ranges respectively. These parameters are less localized in the plunge case, yielding median values of \credval{E_0/M}{1.0550}{0.0438}{0.0164} and \credval{p_\phi^0}{3.7140}{0.778}{1.086}, spanning $\sim$30\% and $\sim$21\% of the $E_0/M$ and $p_\phi^0$ prior ranges respectively. Note that the mass ratio is the most degenerate, yielding median values of \credval{q}{0.8188}{0.1574}{0.2586} and \credval{q}{0.8113}{0.1679}{0.2528} for the scatter and plunge cases respectively, spanning $\sim$46\% of the $q$ prior range in both cases. One can see that we also recover the injected zero-spin, yielding median values of \credval{\chi_{\rm eff}}{0.0043}{0.0453}{0.0335} and \credval{\chi_{\rm eff}}{0.0203}{0.2728}{0.2444} for the scatter and plunge cases, spanning $\sim$4\% and $\sim$26\% of the $\chi_{\rm eff}$ prior range respectively.\par



\begin{figure}
    \centering
    \includegraphics[width=7cm]{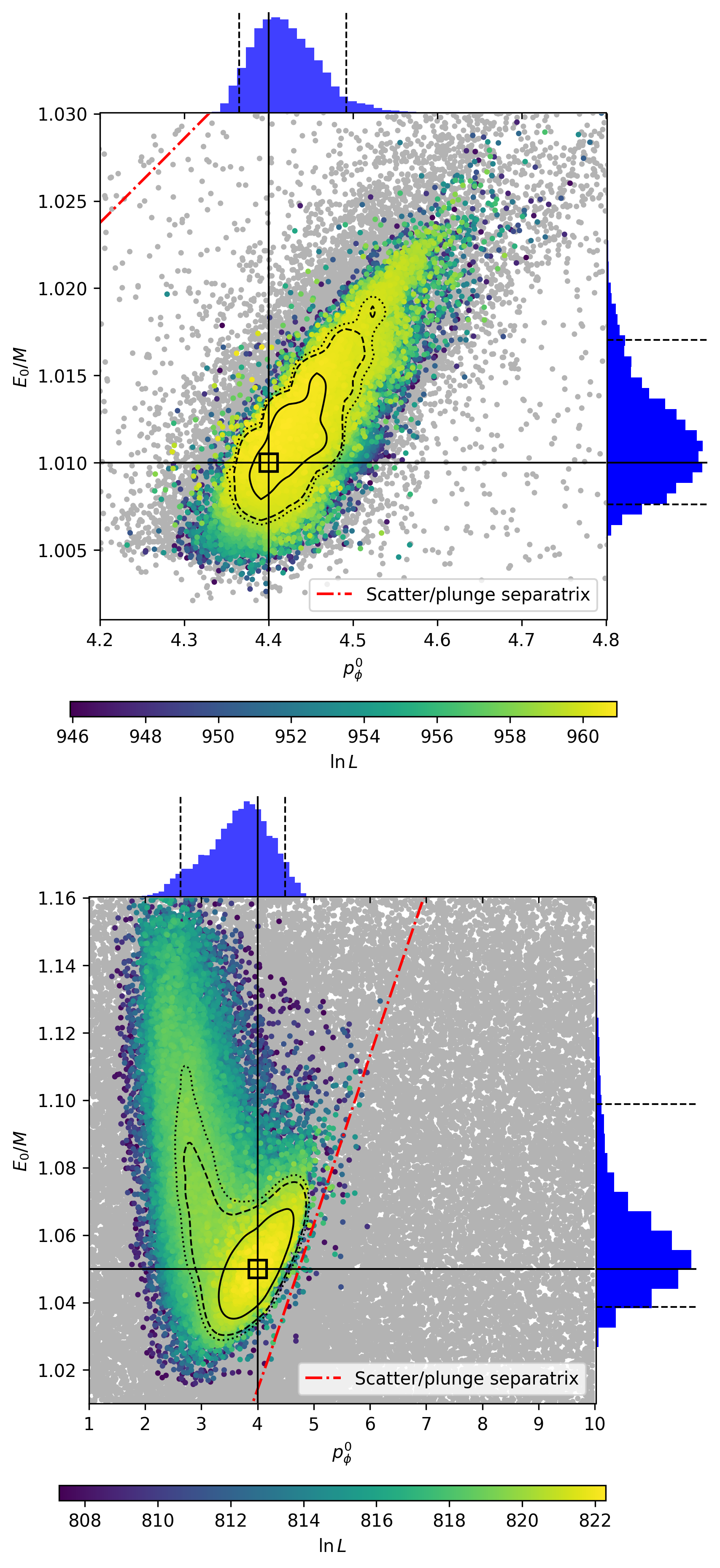}
    \caption{Similar plots to the $E_0 / M, p_\phi^0$ panels in Fig.(\ref{combined_corner_free}), this figure shows parameter estimation results for simulated scatter (top panel) and plunge (bottom panel) signals with zero added noise at SNR$\sim 42$ from a single-detector network at the design sensitivity of Cosmic Explorer~\cite{CE2019}. The 1-D histograms display posterior probability distributions for the parameters $E_0 / M, p_\phi^0$ and their $5\%$ and $95\%$ percentiles, while the 2-D plot displays the distribution of marginalized likelihood in the joint parameter space and contours enclosing the $68\%$, $90\%$, and $95\%$ credible regions. The red dash-dot line shows the separatrix between the scatter and plunge regions of the parameter space. As stated in the text, the SNR for both the scatter and plunge event is ~ 42.}
    \label{combined_2d}
\end{figure}

To get a better understanding of these unique systems, we focus on the hyperbolic parameters in Fig.(\ref{combined_2d}), where the top panel shows the scatter simulation and the bottom panel shows the plunge simulation\footnote{These are similar plots to the $E_0 / M, p_\phi^0$ 2-D corner plots from Fig.( \ref{combined_corner_free}).}. In this diagram the 1-D posteriors are shown on the top and right-hand axes, with the black crosshairs denoting the injected parameters. The dashed black lines on the 1D posteriors denote the the $5\%$ and $95\%$ percentiles, and the contour lines on the 2-D distribution denote the $68\%$, $90\%$, and $95\%$ credible intervals. We note interesting features from the plunge case; for one, there appears to be a hard cutoff in the marginalized likelihood distribution around $p_\phi^0 \approx 4.6$. This is not the ``railing" commonly seen in PE due to lack of exploration, but a natural physical boundary in the hyperbolic parameter space where the waveform generation transitions from plunge to scatter~\cite{Nagar2020} - we denote this boundary with the red separatrix line. Note that for the scatter case the injected value is far enough away from the separatrix that this phenomenon is not seen.\par

We also see that in the plunge case the high likelihood region of the system energy extends much higher than for the scatter case; this is likely due to a natural degeneracy in plunge scenarios where higher energy values have minimal effect on the waveform. The increased energy predominantly affects the pre-merger amplitude peak, but the degree to which this information is encoded in the ringdown is unclear; we speculate that the inclusion of higher-order modes in the waveform evaluation could help to break this degeneracy. This likely also applies to the mass ratio, which has a similar effect on the waveform and for CBC cases is strongly encoded in the higher-order modes~\cite{Berti2007, Ohme2013, Kalaghatgi2020, Mills2021,Leong2023}. We also note that past work~\cite{CalderonBustillo2021} has shown that the $\left(l, m\right) = \left(2, 0\right)$ mode can be important in scattering systems, which is omitted here in both this simulated signals as well as the recovered waveforms. We plan to investigate this further in follow-up work.\par


\textit{Closing remarks} - These results demonstrate that parameter estimation of generic hyperbolic waveforms is now possible with RIFT using the \textit{TEOBResumSDALI} model, offering the first comprehensive infrastructure for parameter estimation of such systems. The parameter space for these events is very diverse and degenerate, spanning three different waveform classes that must all be accounted for, requiring classification and conditioning. Herein we have demonstrated examples of the scatter and plunge scenarios, although evaluation of dynamical captures is also possible. As demonstrated, this model is also capable of sampling over aligned spins - in preliminary testing we have found that the aligned spin can be highly degenerate, corroborating the findings of \cite{Fontbute2025}. However with a large enough SNR, we can get good recovery on spin parameters as shown in Fig.(\ref{combined_corner_free}). The \textit{TEOBResumSDALI} model is also able to sample over tidal deformities, opening the possibility in future work of analyzing \bns{} or \nsbh{} hyperbolic systems which could have electromagnetic counterparts~\cite{Tsang2013}, as well as fully precessing spinning systems. \par

We also note that although we demonstrate here using the projected sensitivities of Cosmic Explorer, it is also possible to get comparable results at the design sensitivity of Advanced LIGO, but may require very high SNR of $85+$. Our current speculation is that this SNR discrepancy is due to the greatly improved sensitivity of Cosmic Explorer in the $5 - 10\,$ Hz band. A full injection study covering this parameter space is currently underway, and will be described along with the technical details of our implementation in RIFT in a forthcoming publication.\par

\section*{Acknowledgments}

We thank Rossella Gamba and Danilo Chiaramello for helpful discussions and support with the \textit{TEOBResumSDALI} waveform model, and James Clark for technical assistance in code deployment on the International Gravitational-Wave Observatory Network (IGWN) pool of resources. The authors are grateful for computational resources provided by the LIGO Laboratory. This material is based upon work supported by NSF's LIGO Laboratory which is a major facility fully funded by the National Science Foundation. LIGO Laboratory and Advanced LIGO are funded by the United States National Science Foundation (NSF) as well as the Science and Technology Facilities Council (STFC) of the United Kingdom, the Max-Planck-Society (MPS), and the State of Niedersachsen/Germany for support of the construction of Advanced LIGO and construction and operation of the GEO600 detector. Additional support for Advanced LIGO was provided by the Australian Research Council. Virgo is funded, through the European Gravitational Observatory (EGO), by the French Centre National de Recherche Scientifique (CNRS), the Italian Istituto Nazionale di Fisica Nucleare (INFN) and the Dutch Nikhef, with contributions by institutions from Belgium, Germany, Greece, Hungary, Ireland, Japan, Monaco, Poland, Portugal, Spain. KAGRA is supported by the Ministry of Education, Culture, Sports, Science, and Technology (MEXT) in Japan, and is hosted by the Institute for Cosmic Ray Research (ICRR), the University of Tokyo, and co-hosted by High Energy Accelerator Research Organization (KEK) and the National Astronomical Observatory of Japan (NAOJ). This work was supported by NSF grants PHY-2110481, PHY-2409714, PHY-2012057, and PHY-2309172. JL acknowledges support from NSF Grants No. PHY-2207780 and No.
PHY-2114581 as well as support from the Italian Ministry of University and Research
(MUR) via the PRIN 2022ZHYFA2, GRavitational wavEform models for coalescing
compAct binaries with eccenTricity (GREAT). The authors are grateful for computational resources provided by the LIGO Laboratory and supported by National Science Foundation Grants PHY-0757058 and PHY-0823459. 

\centering
\noindent\rule{8cm}{0.4pt}
\section*{References \label{refs}}
\printbibliography[heading=none]


\end{document}